\documentclass[sigconf]{acmart}
\usepackage{amsmath}

\usepackage{graphicx}
\usepackage{xcolor}

\usepackage[caption=false,font=footnotesize]{subfig}

\usepackage{booktabs}
\usepackage{multirow}
\usepackage{makecell}
\usepackage{diagbox}

\usepackage{algorithm}
\usepackage{algorithmicx}
\usepackage{algpseudocode}

\usepackage{tikz}

\usepackage{enumitem}
\setlist{nolistsep}

\usepackage{url}
\usepackage{balance}
\usepackage{afterpage}
\usepackage[most]{tcolorbox}

\AtBeginDocument{%
  }

\newtcolorbox{mybox}{
enhanced,
boxrule=0pt,frame hidden,
borderline west={4pt}{0pt}{gray!75!black},
colback=gray!10!white,
sharp corners
}

\newcommand{\circled}[1]{%
\tikz[baseline=(n.base)]{
\node[circle, fill=black, inner sep=1.5pt] (n)
{\textcolor{white}{\footnotesize #1}};
}}

\begin{document}

\title{EcoShift: Performance-Aware Power Management for Power-Constrained Heterogeneous Systems}

\begin{abstract}
Power-constrained HPC systems increasingly run heterogeneous CPU--GPU applications under strict cluster-wide power limits. Existing cluster-wide power management policies rely on fair-share or utilization heuristics and do not capture application-specific sensitivity to CPU and GPU power caps, leading to inefficient use of reclaimed power.

We present EcoShift, a performance-aware cluster-wide power management framework. EcoShift combines online performance prediction with a dynamic-programming-based allocator to distribute reclaimed power across CPU--GPU applications for maximum average performance improvement.

Through emulation-based evaluation on two heterogeneous Intel CPU and NVIDIA A100/H100 GPU platforms with diverse CPU--GPU workloads, EcoShift consistently outperforms state-of-the-art policies, achieving up to 6\% average performance improvement while preserving the cluster-wide power constraint.
\end{abstract}




\author{Zhong Zheng}
\affiliation{
  \institution{Department of Computer Science}
  \institution{University of Illinois Chicago}
 \city{Chicago}
 \country{USA}
}
\email{zzheng33@uic.edu}

\author{Michael E. Papka}
\affiliation{
  \institution{Argonne National Laboratory}
  \institution{University of Illinois Chicago}
   \city{Lemont}
   \country{USA}
}
\email{papka@anl.edu}

\author{Zhiling Lan}
\affiliation{
  \institution{University of Illinois Chicago}
  \institution{Argonne National Laboratory}
   \city{Chicago}
 \country{USA}
}
\email{zlan@uic.edu}

\maketitle

\begin{figure*}
    \centering
    \includegraphics[width=0.85\linewidth]{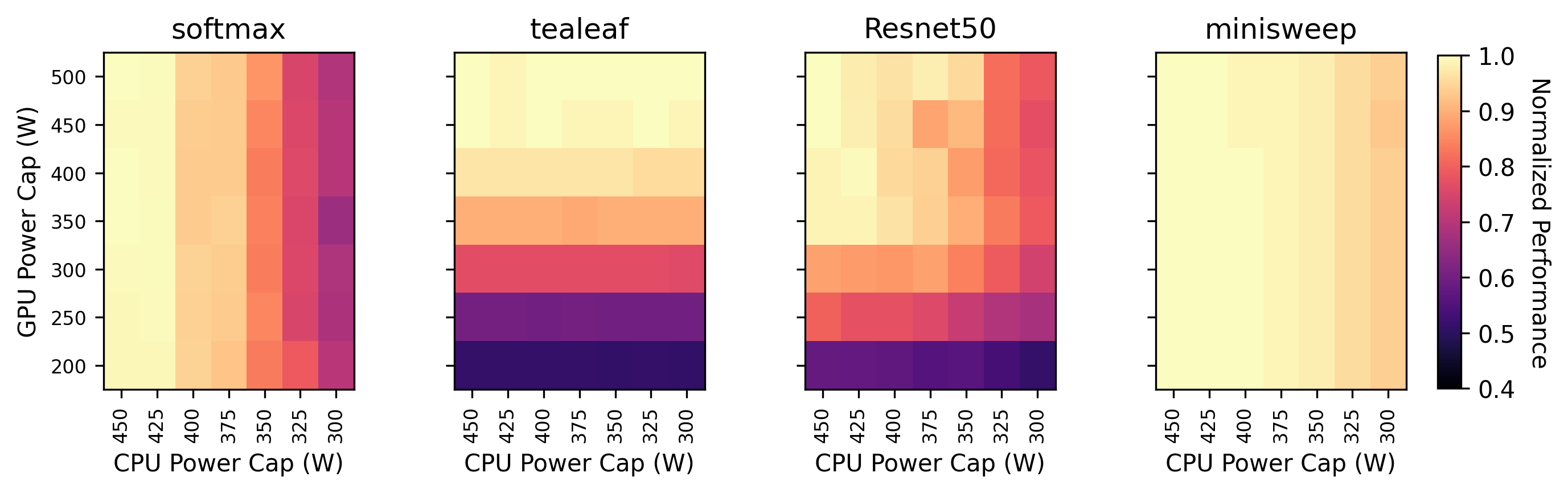}
    \caption{Heatmaps of normalized application performance on a node with an Intel Xeon Platinum 8468 CPU and NVIDIA H100 GPU under varying CPU and GPU power caps. Four representative cases illustrate distinct sensitivity profiles across applications.}
    \label{fig:dual_power_cap}
\end{figure*}

\section{Introduction}
High-performance computing (HPC) drives progress across many scientific and engineering domains by enabling the solution of complex problems using large-scale computation, memory, storage, and networking resources. However, modern HPC systems face increasingly tight energy constraints as they approach the exascale era. Cluster-wide power capping has therefore been proposed as a practical approach, where systems are designed to draw power beyond the nominal budget while enforcing node-level power limits to respect a cluster-wide power constraint \cite{patki2013exploring}. Mechanisms such as \emph{power capping} \cite{lefurgy2008power}, and \emph{dynamic voltage and frequency scaling (DVFS)} \cite{dvfs} are widely used to enforce these limits, using interfaces such as Intel RAPL \cite{david2010rapl} and NVIDIA NVML \cite{NVML}. 

In production HPC systems, power consumption varies significantly across applications, and it is rare for all workloads to simultaneously reach peak power. For instance, Cornelius et al.~\cite{cornelius2025extracting} analyze the Polaris supercomputer and report that applications consume, on average, only about 25\% of the available GPU power. 

As a result, under a cluster-wide power budget enforced via uniform per-node caps, some applications may be power constrained, while others consistently draw substantially less than their allocated budget due to workload characteristics~\cite{patki2013exploring}. We define the gap between an application’s assigned power budget and its observed power draw as \emph{reclaimed power}. This reclaimed power can potentially be pooled and redistributed to other applications to improve their performance without violating the cluster-wide power constraint. However, uniform distribution policies often break down because applications exhibit distinct power--performance relationships: the same additional power budget can yield negligible improvement for one application while significantly accelerating another \cite{zheng2025coordinated}.

Modern workloads increasingly leverage both CPUs and GPUs (hereafter \emph{heterogeneous applications}, in contrast to \emph{CPU-only applications}), including GPU-accelerated scientific codes and deep learning workloads. These applications are highly diverse in their resource usage: some are predominantly GPU-bound, while others remain CPU-bound due to control flow, synchronization, or frequent CPU--GPU data transfers \cite{hong2010integrated, lee2011improving}. This diversity complicates cluster-wide power distribution because effective decisions must account for each application’s asymmetric sensitivity to CPU and GPU power caps and the diminishing returns of extra power.

Although several prior studies have explored cluster-wide power distribution of reclaimed power \cite{ding2023dps, wilson2021introducing, srivastava2022penelope, yoo2003slurm}, most approaches either redistribute reclaimed power in a fair-share order \cite{yoo2003slurm, ding2023dps} or allocate power proportionally based on estimated demand \cite{wilson2021introducing}. These heuristics do not explicitly model application-specific marginal performance gains per additional watt across both CPU and GPU, which can lead to suboptimal allocations for heterogeneous workloads.

To address this limitation, we present \textbf{EcoShift}, a performance-aware cluster-wide power distribution framework for heterogeneous CPU--GPU workloads. EcoShift takes a reclaimed-power budget as input and allocates it to the applications that can convert that budget into the largest relative performance gains under the current CPU and GPU cap levels. Throughout the paper, we use runtime as the performance measure (lower is better), and EcoShift's optimization target is the average relative runtime reduction over receiver applications under a fixed reclaimed-power budget. The central question we study is:

\begin{mybox}
In a power-constrained heterogeneous computing environment that executes diverse applications, how can we distribute reclaimed power across CPU--GPU workloads to maximize average performance improvement under a fixed reclaimed-power budget?
\end{mybox}

Answering this question requires overcoming two challenges. First, the system must characterize each application's performance surface across CPU--GPU cap pairs \emph{online}, since production HPC platforms execute a continual stream of diverse workloads whose power--performance behavior is not known in advance. Second, the system must compute a high-quality cluster-wide allocation quickly enough for online use.

EcoShift addresses these challenges by combining two components: an online performance predictor \cite{zheng2025coordinated} that estimates application performance over a wide range of CPU--GPU cap pairs, and a lightweight dynamic-programming-based allocator that selects a near-optimal distribution of reclaimed power. The dynamic program itself is a standard solver for the discretized multiple-choice knapsack formulation; EcoShift's contribution is to make this formulation practical for cluster-wide heterogeneous power management by coupling it with online-learned CPU--GPU performance surfaces.

We implement EcoShift as an open-source system [link\footnote{GitHub link will be provided upon acceptance}] and evaluate it through emulation-based policy studies using a diverse set of heterogeneous CPU--GPU workloads on two platforms equipped with Intel CPUs and NVIDIA A100 and H100 GPUs (\S\ref{sec:config}). The results show that EcoShift predicts performance under diverse CPU--GPU cap settings with mean accuracy of 93--95\% and consistently outperforms state-of-the-art cluster-wide power management policies by up to 6\% in average performance improvement.

Overall, this work offers the following key contributions:

\begin{itemize}
    \item We analyze the impact of CPU and GPU power capping on heterogeneous applications and characterize the resulting performance and power behavior (\S\ref{sec:motivation}).
    \item We develop EcoShift, a cluster-wide power distribution framework that allocates reclaimed power budgets to maximize average performance improvement. EcoShift employs a dynamic-programming-based search algorithm to rapidly identify near-optimal power budget distributions (\S\ref{design}--\S\ref{implementation}).
    \item We extensively evaluate EcoShift on heterogeneous systems using a broad set of benchmarks and applications (\S\ref{sec:config}--\S\ref{result}).
\end{itemize}

\section{Motivation and Challenges} \label{sec:motivation}

Heterogeneous CPU--GPU applications exhibit diverse and often asymmetric sensitivity to CPU and GPU power caps. As a result, cluster-wide policies that distribute reclaimed power using fair-share heuristics \cite{yoo2003slurm, ding2023dps} or proportional-to-demand rules \cite{wilson2021introducing} can be suboptimal in heterogeneous settings.

To quantify this heterogeneity, we conduct extensive power--performance characterization on a system equipped with Intel Xeon CPUs and NVIDIA H100 GPUs. For each application, we sweep CPU and GPU power caps over a wide range and record runtime. Our key finding is that \emph{power capping one component (CPU or GPU) can be either performance-neutral or performance-critical, depending on the workload}.

Figure~\ref{fig:dual_power_cap} illustrates four representative applications with distinct sensitivity profiles (Table~\ref{tab:apps}). \textit{softmax} is highly sensitive to CPU power capping due to communication and memory bottlenecks, but largely insensitive to GPU caps. \textit{tealeaf} is primarily GPU-bound, making it sensitive to GPU capping but relatively insensitive to CPU caps. \textit{ResNet50} exhibits sensitivity to both CPU and GPU caps, reflecting mixed CPU-side orchestration and GPU computation. In contrast, \textit{minisweep} is largely insensitive to either cap within the evaluated range.
Prior work \cite{zheng2025coordinated} groups heterogeneous applications into coarse sensitivity classes (CPU-sensitive, GPU-sensitive, both-sensitive, and insensitive). However, such coarse labels alone are insufficient for power distribution: 
for example, although \textit{tealeaf} is GPU-sensitive, increasing the GPU cap from 450~W to 500~W yields only marginal improvement.

\begin{figure}
    \centering
    \includegraphics[width=1\linewidth]{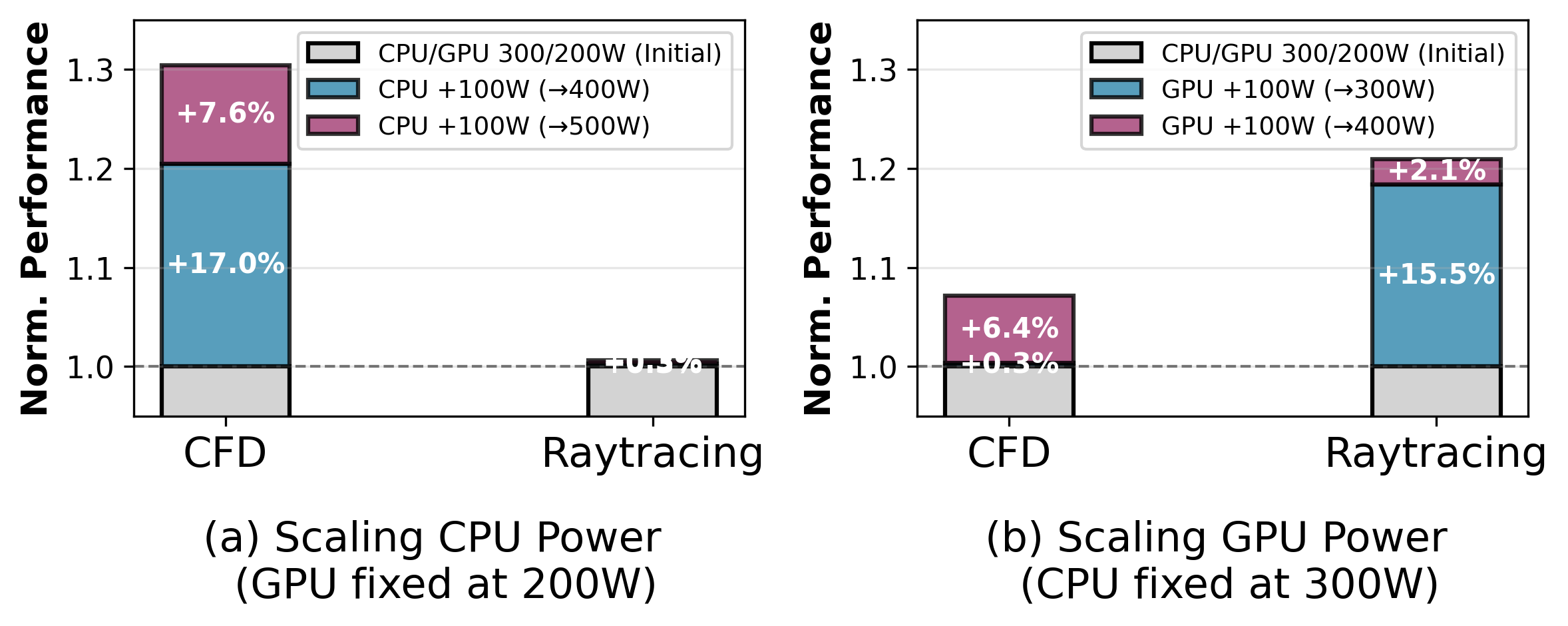}
    \caption{Incremental performance gain from extra GPU \& CPU power budget on a node with Intel(R) Xeon(R) Platinum 8468 processors with NVIDIA H100-80GB GPU on two applications from Altis benchmarks. The initial CPU/GPU power caps are set to 300 W and 200 W, respectively. In (a), we fix the GPU power cap and vary the CPU power cap, whereas in (b) we fix the CPU power cap and vary the GPU power cap. Increasing same amount of power budget leads to different performance gains across different applications, processors and the current power cap level.}
    \label{fig:motivation}
    \vspace{-1.5em}
\end{figure}

To further investigate the diminishing returns and cross-component effects, we conduct a sensitivity study (Figure~\ref{fig:motivation}) using a CPU-sensitive application, \textit{cfd}, and a GPU-sensitive application, \textit{raytracing}, on an Intel Xeon Platinum 8468 + NVIDIA H100 node. Starting from CPU/GPU caps of 300~W/200~W, \textit{cfd} improves by 17\% when the CPU cap increases from 300~W to 400~W, but only by 7.6\% when increasing from 400~W to 500~W, demonstrating diminishing marginal returns. Similarly, \textit{raytracing} improves by 15.5\% when the GPU cap increases from 200~W to 300~W, but only by 2.1\% when increasing from 300~W to 400~W. Finally, additional GPU power provides little benefit to \textit{cfd}, while additional CPU power provides limited benefit to \textit{raytracing}, highlighting the importance of cross-component sensitivity.

Overall, the benefit of reclaimed power depends jointly on three factors: the application's CPU--GPU sensitivity, the current cap pair, and the amount of extra power available. Therefore, a cluster-wide policy should not ask only \emph{who has spare power} or \emph{who wants more power}; it should ask \emph{which application can deliver the largest marginal performance gain for the next watt at the current operating point}.

Existing cluster-wide methods do not explicitly answer this question. Prior approaches typically redistribute reclaimed power using fair-share rules \cite{yoo2003slurm, ding2023dps} or proportional-to-demand heuristics \cite{wilson2021introducing}. Such policies can miss the best allocation because they do not model application-specific marginal gains across both CPU and GPU dimensions.

These observations motivate EcoShift. A practical performance-aware policy must solve two problems at once: it must estimate application response to CPU--GPU power changes, and it must search a large combinatorial allocation space quickly enough for online deployment.

Specifically, EcoShift must address two key challenges:
\begin{enumerate}
\item \emph{Online characterization of heterogeneous CPU--GPU applications.} Diverse and previously unseen applications can exhibit complex cross-component interactions under different power caps. Exhaustive offline profiling is costly and impractical in production, so EcoShift must perform fast and accurate online characterization.

\item \emph{Scalability of cluster-wide optimization.} Finding the optimal power distribution across applications to maximize average performance improvement is NP-hard. The combinatorial search space grows rapidly with the number of applications and available power-cap configurations, making brute-force exploration infeasible in practice.
\end{enumerate}

\section{EcoShift Design} \label{design}

The results above show that effective cluster-wide power distribution relies on accurate characterization of application performance under diverse power caps; however, deriving the performance--power relationship for each application remains challenging, particularly for previously unseen workloads without prior extensive offline profiling (\emph{the heterogeneity challenge}). In addition, the increasing scale of applications makes it difficult to find optimized power caps that maximize performance improvement across all applications (\emph{the scalability challenge}).

\begin{figure}
    \centering
    \includegraphics[width=1\linewidth]{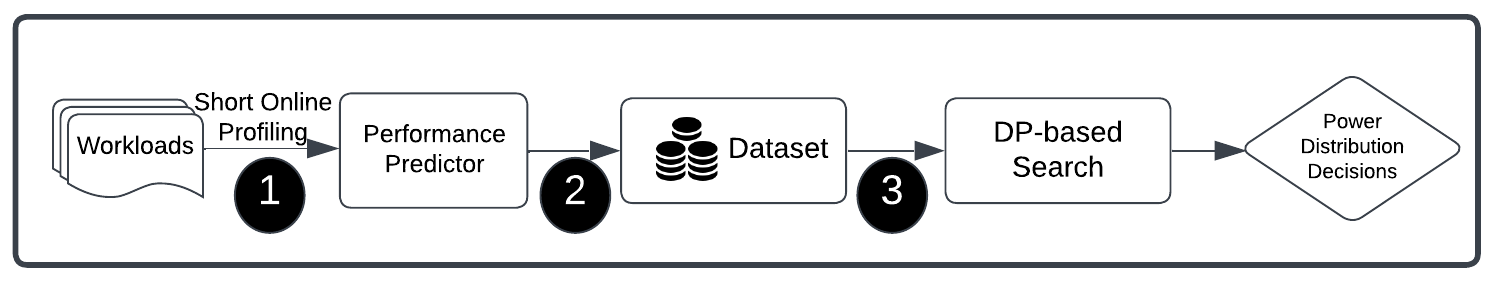}
    \caption{Workflow of EcoShift. EcoShift begins with a brief online profiling phase, during which all applications are monitored concurrently. Based on the collected runtime signals, it predicts application performance under different power-cap configurations. Finally, EcoShift employs a dynamic-programming-based search to determine the optimal cluster-wide power distribution.}
    \vspace{-1.5em}
    \label{fig:workflow}

\end{figure}

Figure~\ref{fig:workflow} shows the EcoShift workflow. EcoShift first performs \circled{1} lightweight online profiling for unseen applications. It then uses the \circled{2} predictor from Zheng et al.~\cite{zheng2025coordinated}, which employs neural collaborative filtering to infer performance over a wide range of CPU and GPU cap pairs from a small number of online samples. This step provides the per-application performance surface needed by the allocator without exhaustive profiling. Finally, EcoShift invokes \circled{3} a dynamic-programming-based search to compute the reclaimed-power allocation across applications. We evaluate the predictor in \S\ref{result}; its mean accuracy is approximately 95\% on one system and 93\% on the other.

\subsection{Performance Prediction}
\label{sec:performance_prediction}

EcoShift relies on an online performance predictor \cite{zheng2025coordinated} to estimate each application’s performance (runtime) under different CPU--GPU power cap pairs. The performance predictor treats performance prediction as a matrix-completion problem, where rows correspond to applications and columns correspond to CPU--GPU power-cap configurations; only a small subset of entries can be measured online due to profiling cost. To infer the unmeasured entries, it uses a neural collaborative filtering (NCF) model that learns latent embeddings for applications and power-cap configurations and predicts performance from their interaction.

At runtime, EcoShift performs lightweight online profiling by sampling a few representative CPU--GPU cap pairs for an unseen application and measuring its performance for a short period. These samples are then used to (i) identify the new application’s embedding (and optionally fine-tune the model) and (ii) predict the full performance surface over the feasible cap grid. This design avoids exhaustive offline profiling while still capturing the heterogeneous and cross-component (CPU--GPU) performance response to power capping, enabling EcoShift to quickly evaluate candidate power reallocations during optimization.

\begin{figure}
    \centering
    \includegraphics[width=1\linewidth]{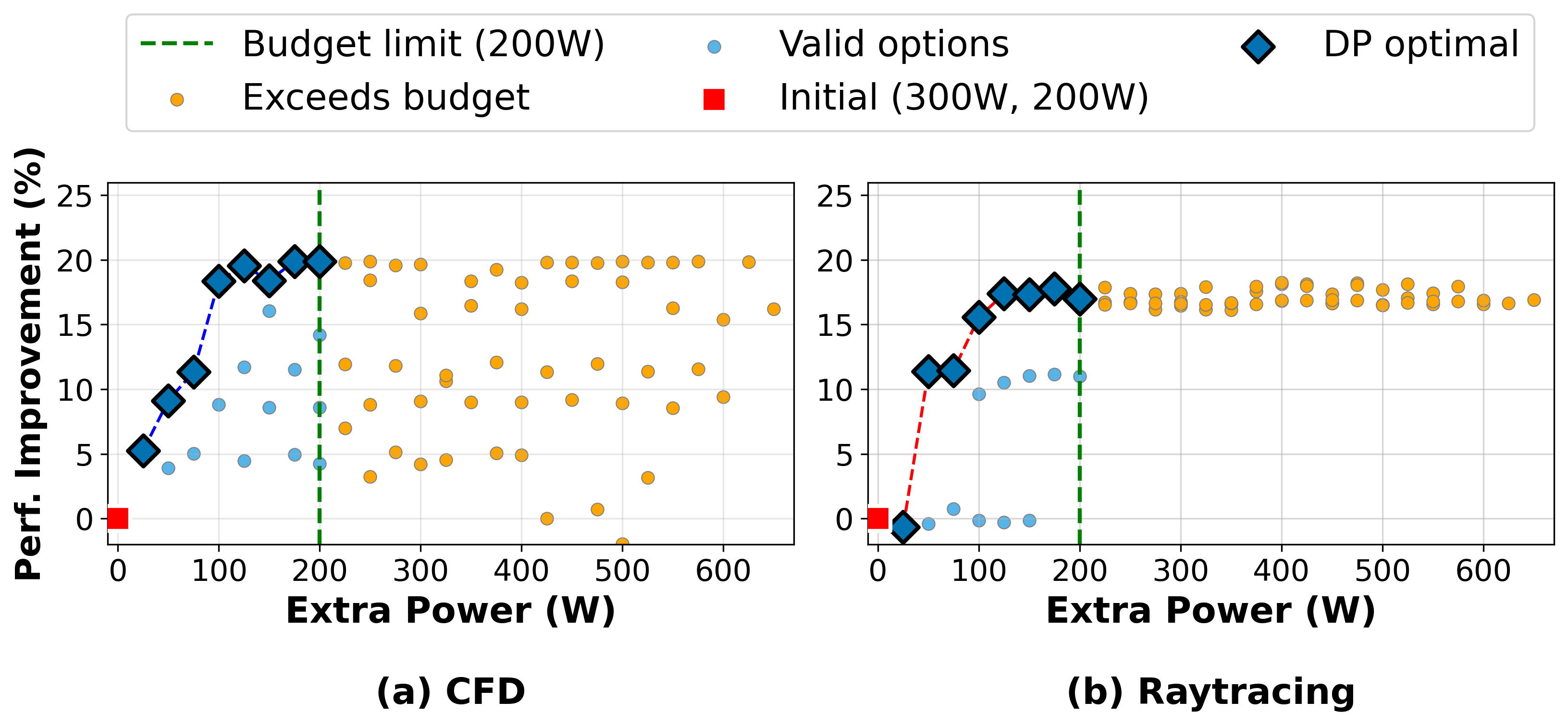}
    \caption{Dynamic Programming Search Space. Given a fixed node power budget, EcoShift determines the optimal extra CPU and GPU power distribution for an application, producing an optimal performance curve that serves as the input to the DP algorithm.}
    \label{fig:dp_search}
    \vspace{-1.5em}
\end{figure}

\subsection{Optimal Power Distribution Search}

We consider a cluster running $M$ heterogeneous CPU--GPU applications under a fixed cluster-wide power budget $W$. The budget is initially distributed evenly across applications. Some applications draw less than their assigned caps and therefore create a reclaimed-power pool of size $B$; others can benefit from additional power. Let $\mathcal{A}$ denote the set of running applications with $|\mathcal{A}|=M$, and partition it into donors $\mathcal{D}$ and receivers $\mathcal{R}$, where $\mathcal{D}\cap\mathcal{R}=\emptyset$ and $\mathcal{D}\cup\mathcal{R}=\mathcal{A}$. Let $N \triangleq |\mathcal{R}|$ denote the number of receivers. EcoShift optimizes power allocation only over the applications in $\mathcal{R}$, subject to the reclaimed-power budget $B$ supplied by donors. Our focus is the \emph{distribution} problem: given a reclaimed-power budget, how should it be allocated to maximize average relative performance improvement over receiver applications? The mechanisms used to identify donors and determine $B$ are orthogonal to this design.

\subsubsection{\textbf{Multiple-Choice Knapsack Problem Formulation}}
\label{sec:problem_formulation}

Suppose receiver application $i$ starts from an initial CPU--GPU cap pair $(\bar c_i,\bar g_i)$. Using the predicted performance surface, EcoShift enumerates a finite set $\mathcal{S}_i$ of feasible upgraded cap pairs for application $i$, where each candidate $(c,g)\in\mathcal{S}_i$ satisfies $c \ge \bar c_i$ and $g \ge \bar g_i$. We use runtime as the performance measure, so lower is better. Let $T_i(c,g)$ denote the runtime of application $i$ under caps $(c,g)$. The improvement obtained by selecting $(c,g)$ relative to the baseline $(\bar c_i,\bar g_i)$ is the relative runtime reduction
$ I_i(c,g) \triangleq \frac{T_i(\bar c_i,\bar g_i) - T_i(c,g)}{T_i(\bar c_i,\bar g_i)}$.

The problem can be transformed into the following multiple-choice knapsack problem. Let $x_{i,(c,g)}$ be a binary decision variable that equals 1 if application $i$ is assigned cap pair $(c,g)\in\mathcal{S}_i$ and 0 otherwise. Formally, the optimization is:
\begin{align*}
\max_{x}\quad & \frac{1}{N}\sum_{i=1}^{N}\sum_{(c,g)\in\mathcal{S}_i} I_i(c,g)\,x_{i,(c,g)} \\
\text{s.t.}\quad & \sum_{(c,g)\in\mathcal{S}_i} x_{i,(c,g)} = 1, \quad \forall i, \\
& \sum_{i=1}^{N}\sum_{(c,g)\in\mathcal{S}_i}\big((c-\bar c_i) + (g-\bar g_i)\big)\,x_{i,(c,g)} \le B, \\
& x_{i,(c,g)} \in \{0,1\}, \quad \forall i,\; (c,g)\in\mathcal{S}_i.
\end{align*}

This formulation corresponds to a multiple-choice knapsack problem and is NP-hard. Since the search space grows exponentially with the number of applications and possible power cap choices, exhaustive brute-force search is infeasible for online deployment, motivating efficient approximation algorithms.

\begin{algorithm}
\caption{EcoShift's DP-based Search: Maximize Average Relative Performance Improvement}
\label{alg:ecoshift}
\begin{algorithmic}[1]
\Require Receiver applications $\mathcal{R}$, per-application baseline caps $\{(\bar c_a, \bar g_a)\}_{a\in\mathcal{R}}$, budget $B$, power grid $\mathcal{G}$, improvement estimator $\Delta(\cdot)$
\Ensure Power allocation per app (may be 0 extra power)
\State $\text{options} \leftarrow []$
\For{each app $a \in \mathcal{R}$}
    \State $\text{best}[0] \leftarrow (0,\; (\bar c_a, \bar g_a))$
    \For{each $(P_{cpu}, P_{gpu}) \in \mathcal{G}$}
        \If{$P_{cpu} < \bar c_a$ or $P_{gpu} < \bar g_a$} \State \textbf{continue} \EndIf
        \State $e \leftarrow (P_{cpu}-\bar c_a) + (P_{gpu}-\bar g_a)$
        \If{$e > B$} \State \textbf{continue} \EndIf
        \State $\delta \leftarrow \Delta\big(a, (\bar c_a,\bar g_a), (P_{cpu},P_{gpu})\big)$
        \If{$e \notin \text{best}$ or $\delta > \text{best}[e].\text{impr}$}
            \State $\text{best}[e] \leftarrow (\delta,\; (P_{cpu}, P_{gpu}))$
        \EndIf
    \EndFor
    \State $\text{options}.append(\text{best})$
\EndFor

\State $DP \leftarrow \{0 \mapsto (0,\; \emptyset)\}$
\For{each app $a$ with option table $\text{best}$}
    \State $DP' \leftarrow \emptyset$
    \For{each used power $u$ in $DP$}
        \For{each extra power $e$ in $\text{best}$}
            \If{$u + e > B$} \State \textbf{continue} \EndIf
            \State $s \leftarrow DP[u].\text{impr} + \text{best}[e].\text{impr}$
            \If{$(u+e) \notin DP'$ or $s > DP'[u+e].\text{impr}$}
                \State $DP'[u+e] \leftarrow (s,\; DP[u].\text{alloc} \cup \{a \mapsto \text{best}[e].\text{power}\})$
            \EndIf
        \EndFor
    \EndFor
    \State $DP \leftarrow DP'$
\EndFor

\State $(u^\star, \text{alloc}^\star) \leftarrow \arg\max_{u} DP[u].\text{impr}$
\Return $\text{alloc}^\star$
\end{algorithmic}
\end{algorithm}

\subsubsection{\textbf{Dynamic Programming Based Search}}
\label{sec:dp_search}

We solve the allocation problem with a dynamic-programming-based search. The DP is the standard solver for the multiple-choice knapsack formulation in \S\ref{sec:problem_formulation}: each application forms one choice group, and each candidate CPU--GPU cap pair consumes reclaimed power and yields a predicted improvement. To make the search efficient, EcoShift first compresses each application's discrete option set $\mathcal{S}_i$ into a 1D value-versus-budget curve $F_i(b)$, which records the best improvement achievable when allocating exactly $b$ watts of reclaimed power to application $i$.

For each application $i$, EcoShift computes an application-level improvement function $F_i(b)$, which represents the maximum achievable performance improvement when allocating $b$ watts of reclaimed power:
\begin{equation}
F_i(b) =
\max_{(c,g)\in\mathcal{S}_i \;:\; (c-\bar c_i)+(g-\bar g_i)\le b}
I_i(c,g) 
\end{equation}

At the cluster level, EcoShift allocates reclaimed power across applications using a DP recurrence:
\begin{equation}
DP[i][b] =
\max_{0 \le k \le b}
\bigl(DP[i-1][b-k] + F_i(k)\bigr),
\end{equation}

where $DP[i][b]$ denotes the maximum aggregate relative performance improvement achievable using the first $i$ receiver applications and $b$ watts of reclaimed power. An overview of the algorithm is presented in Algorithm~1.

Figure~\ref{fig:dp_search} illustrates this process for \textit{cfd} and \textit{raytracing}. For each application, EcoShift enumerates feasible CPU--GPU cap pairs under the available extra-power budget and discards infeasible or dominated configurations. It then keeps, for each extra-power level, only the cap pair that yields the highest improvement. The resulting monotone curve becomes the input to the cluster-level DP. This preprocessing step greatly reduces the search space while preserving the best allocation choice at each budget level.

\subsubsection{\textbf{Complexity Analysis}}
\label{sec:complexity}

Next, we quantify the time complexity of the DP-based search. Let $N_r$ be the number of receiver applications and $B$ the reclaimed power budget, discretized at 1~W granularity. Let $S$ denote the average number of feasible CPU--GPU power configurations per receiver application. Constructing the per-application improvement function $F_i(b)$ requires $\mathcal{O}(S \cdot B)$ time. Across all receiver applications, this step incurs $\mathcal{O}(N_r \cdot S \cdot B)$ time.

At the cluster level, the DP recurrence above maximizes over $k \in [0,b]$ for each state $(i,b)$, yielding a worst-case time complexity of $\mathcal{O}(N_r \cdot B^2)$. Equivalently, when expressed as a multiple-choice knapsack over per-application option sets of size $K_i$, the runtime is $\mathcal{O}\bigl(B \cdot \sum_i K_i\bigr)$; if $K_i=\mathcal{O}(B)$ in the worst case, this reduces to $\mathcal{O}(N_r \cdot B^2)$. In practice, the preprocessing step prunes dominated and infeasible configurations and leaves only a small set of distinct extra-power levels per application, so $K_i \ll B$ and the practical runtime scales much more gently than the worst case.

The space complexity is $\mathcal{O}(N_r \cdot B)$, which can be reduced to $\mathcal{O}(B)$ using a rolling-array optimization. In practice, the search overhead is low enough to support online use; we leave a dedicated overhead breakdown to future work.

We also compare the DP-based search with exhaustive brute-force search in \S~\ref{result} using small-scale experiments where brute-force enumeration is tractable. The results show that in over 80\% of the test cases, EcoShift achieves performance improvements within 3\% of the brute-force Oracle solution on both systems.

\section{Implementation}\label{implementation}
We implement EcoShift in Python using native profiling engines that collect runtime signals with minimal disruption, and the code is available as open source on GitHub. Our current prototype targets heterogeneous systems equipped with Intel CPUs and NVIDIA GPUs. 

For GPU telemetry, EcoShift relies on NVIDIA’s Data Center GPU Manager (DCGM)~\cite{DCGM} to sample hardware counters in real time, providing comprehensive visibility without requiring source-code changes. For CPU telemetry, EcoShift employs Linux \emph{perf} to collect hardware performance counters, enabling observation of key processor-level metrics such as memory and instruction throughput. Power measurements are obtained through vendor-provided interfaces, using NVML~\cite{NVML} for GPU power and RAPL (Running Average Power Limit)~\cite{khan2018rapl} for CPU power. Although the current implementation uses vendor-specific monitoring APIs, the overall design is portable, and EcoShift can be deployed on other heterogeneous platforms provided that comparable performance counter and power monitoring interfaces are available.

\begin{table}[htbp]
    \centering
    \caption{Heterogeneous Workload. These workloads span four capping sensitivity classes (§2): CPU cap sensitive (C), GPU cap sensitive (G), sensitive to both (B), and insensitive to both (N).}
    \begin{tabular}{l l l c}
        \toprule
        \textbf{Suite} &  \textbf{App (Class)} & \textbf{Input} \\
        \midrule
        Altis & gemm (C) & -s 4 \\
        & gups (N) & -s 4 \\
        & maxflops (C) & / \\
        & bfs (C) & -s 4 \\
        & particlefilter\_float (G) & -s 4 \\
        & cfd\_double (B) & -s 4 \\
        & particlefilter\_naive (C) & -s 4 \\
        & raytracing (G) & -s 4 \\
        & fdtd2d (G) & -s 4 \\
        & nw (B) & -s 4 \\
        & cfd (C) & -s 4 \\
        & lavamd (C) & -s 4 \\
        & sort (C) & -s 3 \\
        \midrule
        HeCBench & kalman (C) & 10000 10000 10000 \\
        & stencil3d (C) & 1100 \\
        & extrema (B) & default \\
        & knn (C) & default \\
        & dropout (N) & default \\
        & aobench (N) & default \\
        & zoom (C) & 64 32 512 512 \\
        & convolution3D (B) & 32 64 128 56 56 3 \\
        & softmax (C) & 10000 100000 \\
        & chacha20 (N) & default \\
        & zmddft (G) & default \\
        & residualLayerNorm (B) & default \\
        & backgroundSubtract (C) & 1280 2560 2 \\
        \midrule
        MLPerf & UNet (B) & carvana image \\
        & BERT (G) & Wiki \\
        & ResNet50 (B) & ImageNet \\
        \midrule
        ECP Proxy & sw4lite (C) & ps2.in \\
        & XSBench (B) & -s large \\
        & Laghos (N) & box01\_hex.mesh \\
        & miniGAN (B) & bird, 2048 images,  \\
        && 3 channels, 64×64,  \\
        && dim mode 3 \\
        \midrule
        HPC App & GROMACS (C) & Steepest descent,  \\
        & & emtol 1000, PBC xyz,  \\
        && PME, rc 1.0 nm \\
        & LAMMPS (C) & 3D LJ melt, 6.25M, \\
        &&  500 steps, cutoff 2.5 \\
        \midrule
        SPEC & lbm (G) & X 1200, Y 4800\\
        & cloverleaf (C) & X 1500, Y 1500\\
        & tealeaf (G) & X 10000, Y 10000\\
        & minisweep (N) & X=Y=Z 128, NE 16 \\
        && NA 32, NBLOCK\_Z 128 \\
        & pot3d (N) & NR 67 NT 181 \\
        && NP 451\\
        \bottomrule
    \end{tabular}
    \label{tab:apps}
\end{table}

\section{Experimental Setting} \label{sec:config}

\subsection{Comparison Methods} 

In our experiments, we compare EcoShift against two state-of-the-art approaches, MixedAdaptive \cite{wilson2021introducing} and DPS \cite{ding2023dps}, as well as an Oracle baseline. The Oracle assumes perfect knowledge of the power-cap configuration that maximizes average performance improvement across applications. Because finding this configuration requires exhaustive brute-force search over cross-application cap combinations, it quickly becomes computationally infeasible as the number of applications grows. We therefore compare against the Oracle only in the small-scale study in \S\ref{sec:oracle}.

\subsection{Heterogeneous Workloads and Systems}

As presented in Table~\ref{tab:apps}, we assemble a diverse suite of 40 heterogeneous CPU--GPU benchmarks and applications: thirteen from Altis \cite{hu2020altis}, thirteen from HecBench \cite{hec}, three neural-network training workloads from MLPerf \cite{farrell2021mlperf}, four ECP proxy applications \cite{ecp}, two production GPU-enabled molecular-dynamics applications \cite{van2005gromacs, atomic2013lammps}, and five SPEC workloads \cite{dixit1991spec}. These workloads span the four sensitivity categories described in \S2. For evaluation, we organize them into five workload groups: CPU-sensitive, GPU-sensitive, both-sensitive, insensitive, and mixed. The mixed group includes workloads drawn from all sensitivity categories.

Two different heterogeneous systems are used in our evaluation:

\begin{itemize}
    \item \textit{System 1}: two Intel(R) Xeon(R) Platinum 8380 processors paired with NVIDIA A100-40GB GPUs. 
    \item \textit{System 2}: two Intel(R) Xeon(R) Platinum 8468 processors paired with NVIDIA H100-80GB GPUs. 
\end{itemize}

We employ Intel RAPL \cite{david2010rapl} and NVIDIA NVML \cite{NVML} for CPU and GPU power capping, respectively. Our approach is not limited to RAPL or NVML and is compatible with any platform that provides equivalent power capping capabilities.

\subsection{Evaluation Metrics}

EcoShift aims to maximize average performance improvement while also accounting for fairness across applications. Accordingly, we use the following metrics:

\begin{itemize}
    \item \textit{Average Performance Improvement.} We use runtime as the performance measure (lower is better). We quantify improvement as the percentage runtime reduction relative to the no-distribution baseline, and report 98\% confidence intervals.
    
    \item \textit{Fairness.} We measure fairness using Jain’s fairness index \cite{jain} (Equation~\ref{eq:jain}), where $x_i$ denotes the performance improvement of application $i$ and $n$ is the number of applications. The index ranges from $1/n$ to $1$, with higher values indicating more balanced performance improvements across applications.
\end{itemize}

\begin{equation}
\label{eq:jain}
J(x_1, \ldots, x_n)
=
\frac{\left(\sum_{i=1}^{n} x_i\right)^2}
{n \sum_{i=1}^{n} x_i^2}
\end{equation}

\subsection{Emulation-Based Evaluation}
Existing approaches often couple power reclamation and redistribution through fixed or utilization-driven policies \cite{ding2023dps, wilson2021introducing}. EcoShift instead treats reclaimed power as an explicit input and focuses on the distribution problem itself. This separation is important because it lets us evaluate how much benefit can be extracted from a given reclaimed-power budget when application-specific sensitivity and diminishing returns are taken into account.

Because our available hardware cannot realize large cluster-scale experiments directly, we use an emulation-based methodology. EcoShift first predicts each application's performance under different CPU and GPU cap pairs. The DP-based optimizer then determines the cap assignment for a given reclaimed-power budget. Each application is next executed individually under its assigned caps, and the measured runtime reduction is used to compute the resulting average performance improvement. This methodology preserves EcoShift's decision logic while allowing controlled evaluation across many workloads, initial cap settings, and reclaimed-power budgets.

\begin{figure}
    \centering
    \includegraphics[width=1\linewidth]{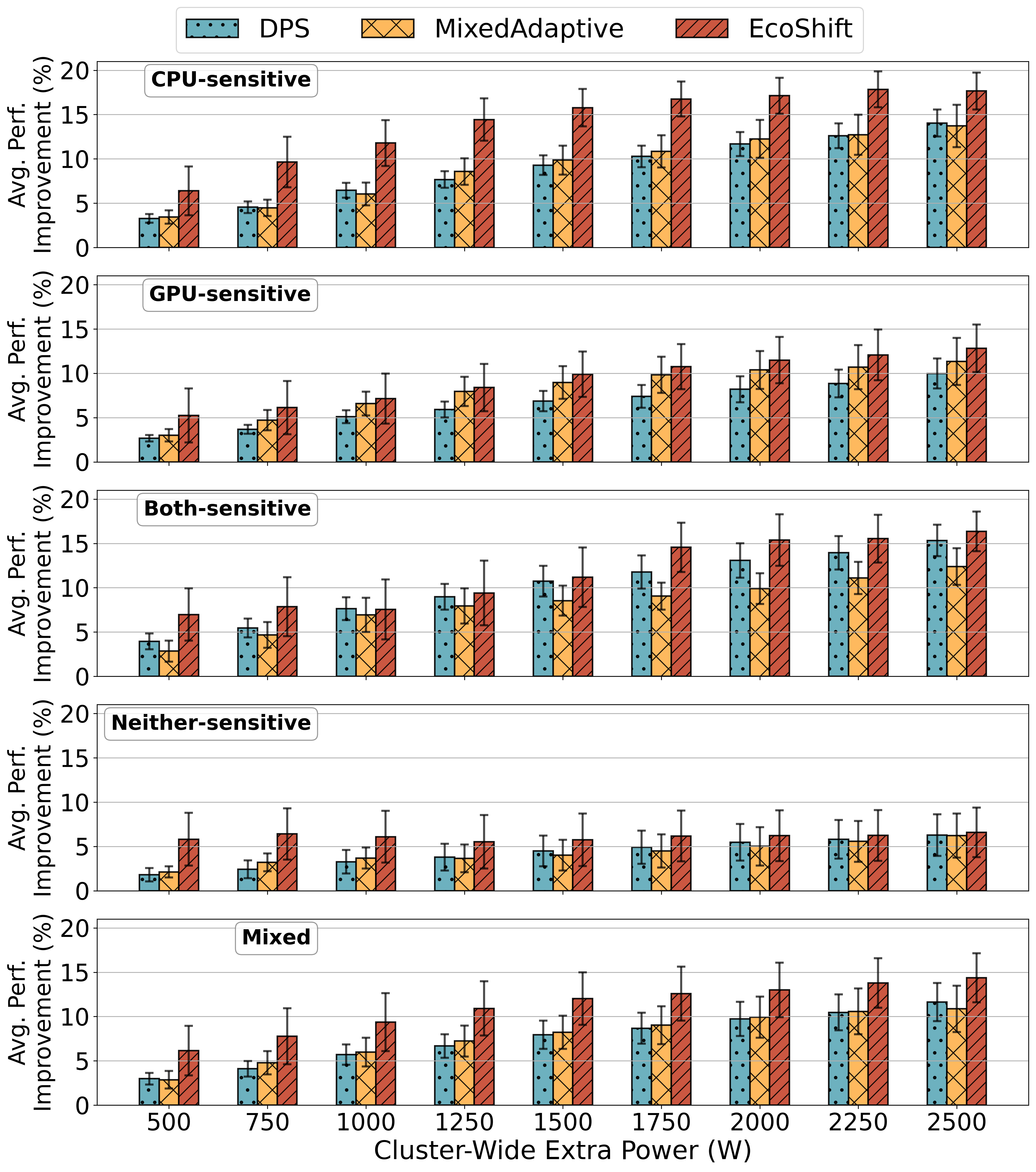}
    \caption{Performance improvement of different power distribution methods across diverse workloads on simulated 100-node System 1 clusters under varying amounts of cluster wide extra power, measured relative to a no distribution baseline The initial power cap is CPU 140 W, and GPU 150 W}
    \label{fig:A100-diff-budget}
    \vspace{-2em}
\end{figure}

\section{Results} \label{result}
Each experiment was repeated five times to account for performance variability and system noise; we report the mean across runs. Our primary objective is to maximize \emph{average performance improvement} when redistributing a fixed amount of reclaimed power.

\begin{figure}
    \centering
    \includegraphics[width=1\linewidth]{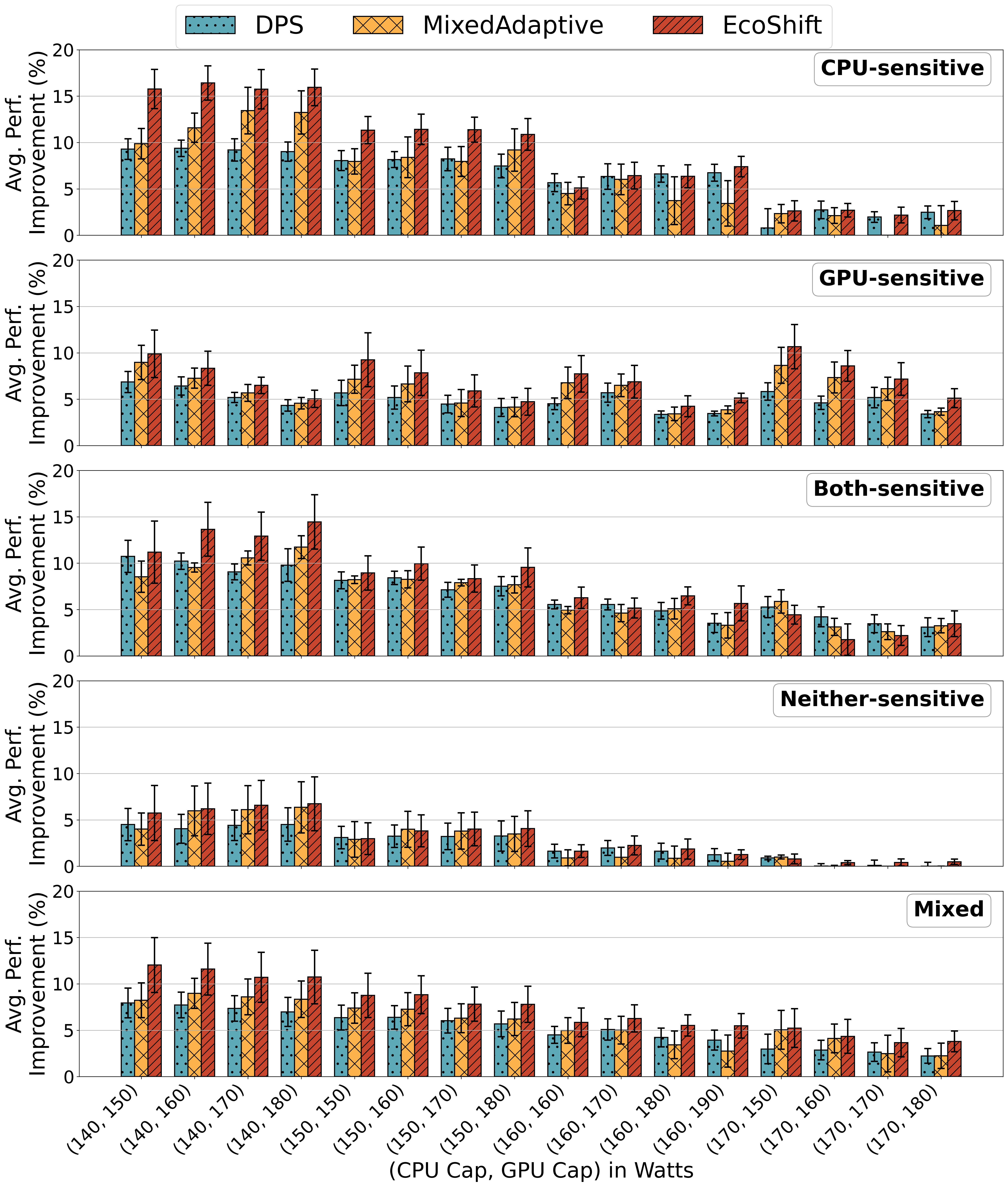}
    \caption{Performance improvement of different power distribution methods across diverse workloads on simulated 100-node System 1 clusters under varying initial CPU and GPU power cap with fixed cluster-wide extra power budget (7000 W), measured relative to a no distribution baseline.}
    \label{fig:A100-diff-cap}
    \vspace{-1.5em}
\end{figure}

\begin{figure}
    \centering
    \includegraphics[width=1\linewidth]{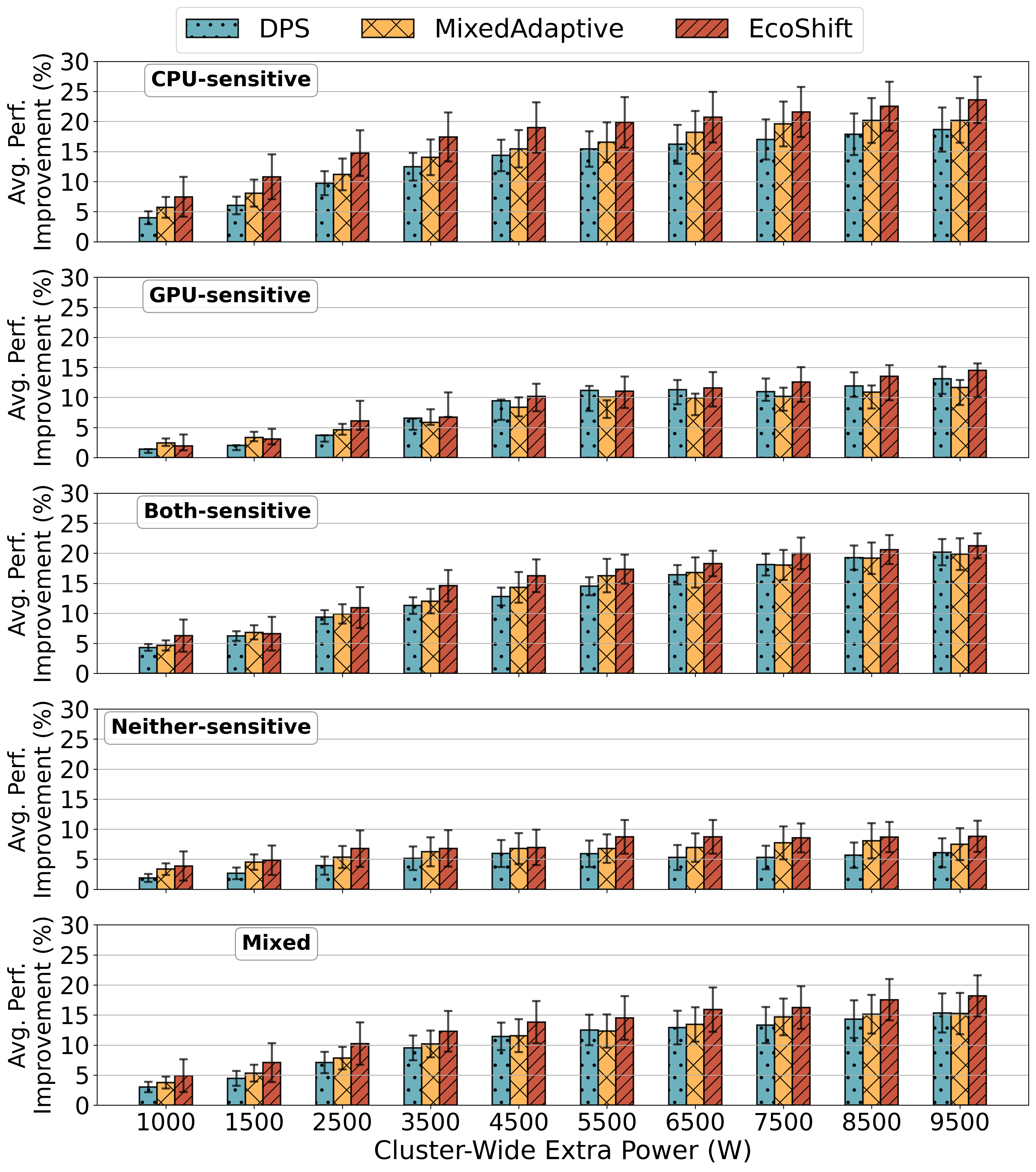}
    \caption{Performance improvement of different power distribution methods across diverse workloads on simulated 100-node System 2 clusters under varying amounts of cluster wide extra power, measured relative to a no distribution baseline. The initial power cap is CPU 300 W, and GPU 300 W}
    \label{fig:H100-diff-budget}
    \vspace{-1.5em}
\end{figure}

\begin{figure}
    \centering
    \includegraphics[width=1\linewidth]{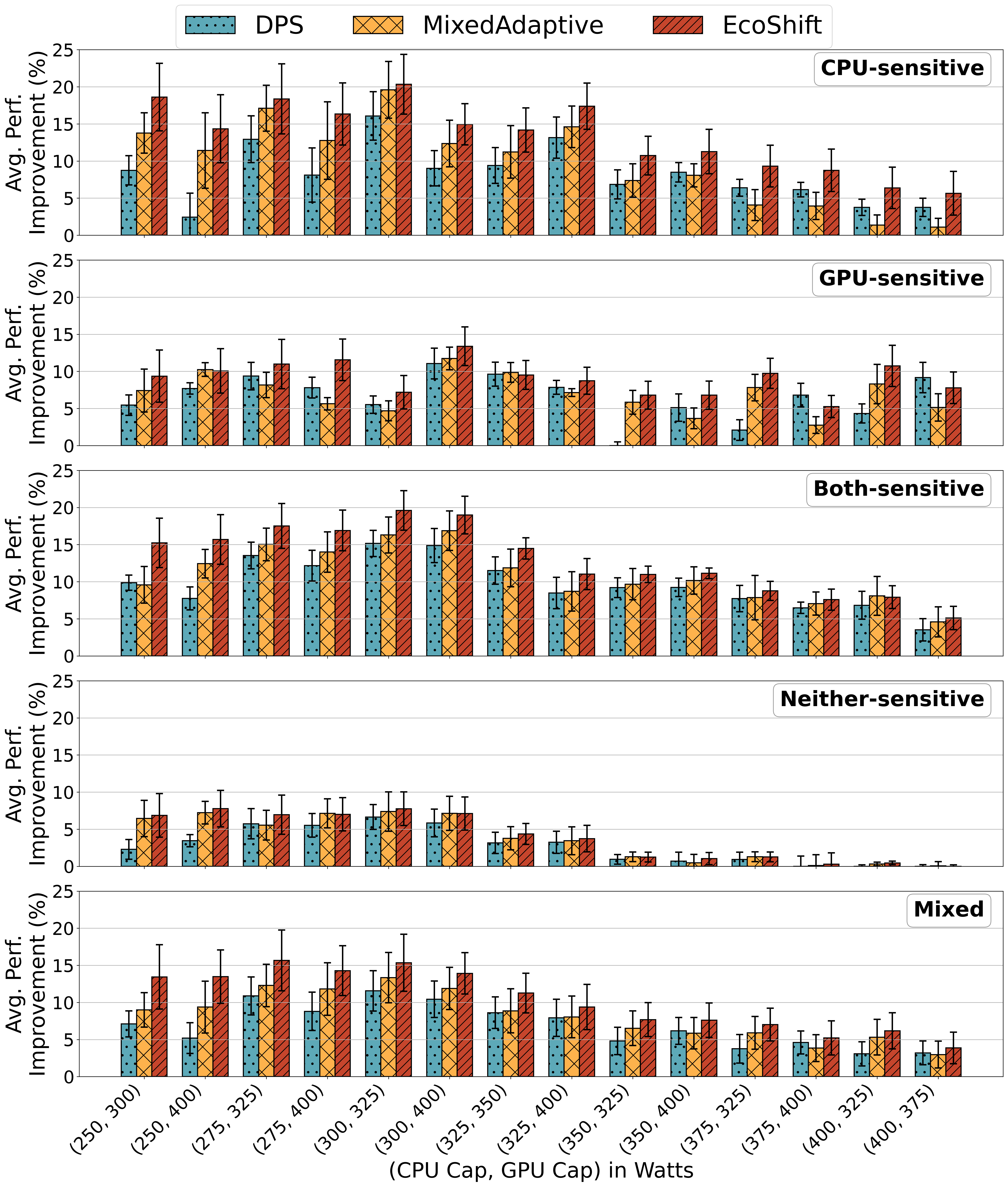}
    \caption{Performance improvement of different power distribution methods across diverse workloads on simulated 100-node System 2 clusters under varying initial CPU and GPU power cap with fixed cluster-wide extra power budget (14000 W), measured relative to a no distribution baseline.}
    \label{fig:H100-diff-cap}
    \vspace{-1.5em}
\end{figure}

\begin{figure*}
    \centering
    \includegraphics[width=0.95\linewidth]{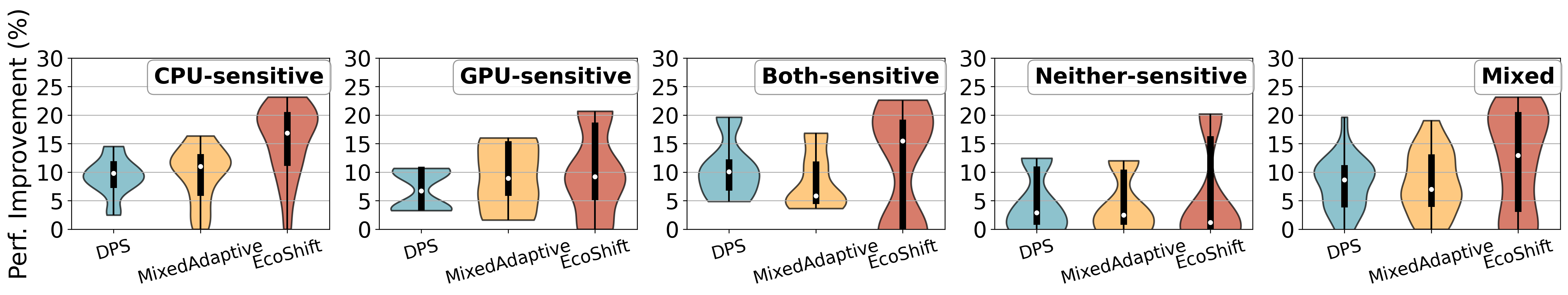}
    \caption{Violin plots showing the distributions of performance improvement across different type of workloads under different power distribution methods. The experiment is emulated on a 100-node system 1 cluster with an initial CPU and GPU power cap of 140 W and 150 W respectively.}
    \label{fig:violin}
\end{figure*}

The following subsections present the emulation-based policy evaluation of EcoShift (Figures~\ref{fig:A100-diff-budget}--\ref{fig:H100-diff-cap} and Figure~\ref{fig:violin}), a detailed case study (Table~\ref{tab:strategy_improvements}), the effectiveness of the DP-based search relative to an Oracle (Figure~\ref{fig:cdf}), and a fairness analysis (Figure~\ref{fig:fairness}).

\subsection{Emulation-Based Policy Evaluation}
We first compare EcoShift with DPS and MixedAdaptive in terms of average performance improvement. As discussed in \S\ref{sec:motivation}, the gain from redistribution depends on three factors: application sensitivity to CPU and GPU caps, the initial cap pair, and the reclaimed-power budget $B$. The experiments below vary these factors in a controlled way.

Figure~\ref{fig:A100-diff-budget} fixes the initial cap pair at 140~W CPU and 150~W GPU and varies $B$. Each row corresponds to one workload category from \S\ref{sec:config}. The x-axis shows the reclaimed power redistributed across applications, and the y-axis shows the resulting average performance improvement relative to the no-distribution baseline. Figure~\ref{fig:A100-diff-cap} instead fixes $B$ at 3500~W and sweeps the initial CPU/GPU cap pairs. This emulation-based experiment tests whether the policy remains effective when the system starts in a tight-cap regime versus a relatively power-sufficient regime. EcoShift's advantage is largest when the initial caps are tight, where there is more room for performance-aware reallocation; as the initial caps increase, all methods converge because the workloads become less power-constrained.

Three trends are clear from the System~1 results. First, across workload categories, reclaimed-power budgets, and initial cap settings, EcoShift consistently matches or exceeds DPS and MixedAdaptive, with improvements of up to 6\%. Second, EcoShift also tends to achieve higher upper bounds in the 98\% confidence intervals, indicating that it can better exploit favorable sensitivity when such opportunities exist. Third, gains remain small for insensitive workloads under all methods, which is expected because these applications respond weakly to additional power. The same qualitative behavior appears on System~2 in Figures~\ref{fig:H100-diff-budget} and \ref{fig:H100-diff-cap}. Thus, the ranking of methods is stable across both platforms and workload groups: EcoShift is consistently comparable to or better than DPS and MixedAdaptive because it allocates reclaimed power jointly across CPU and GPU instead of relying on fixed-share or demand-based heuristics.

Beyond average performance improvement, it is also important to examine the distribution of performance gains across individual applications. Figure~\ref{fig:violin} presents violin plots showing the distributions of application level performance improvement across different workload categories under various power distribution methods. Across most workload types, EcoShift exhibits distributions that are shifted toward higher performance improvements, indicating that EcoShift not only increases average performance but also enables a larger fraction of applications to achieve higher performance gains. 

Because EcoShift’s allocations rely on predicted performance surfaces, we next verify the prediction accuracy of the performance predictor introduced in \cite{zheng2025coordinated}. For each CPU–GPU power cap configuration, prediction accuracy is defined as $\mathrm{Acc}=1-|\hat p-p|/p$, where $p$ and $\hat p$ represent the measured and predicted normalized performance relative to the baseline. We report the mean accuracy across applications and CPU--GPU cap points. On system 2, the predictor achieves a mean accuracy of 93.12\%, with a 98\% confidence interval of [92.64\%, 94.01\%]. Similar accuracy is observed on system 1, where the mean accuracy reaches 95.12\%, with a 98\% confidence interval of [94.64\%, 96.01\%]. Given this \mbox{93--95\%} mean accuracy, residual error primarily affects \emph{close-call} allocations in which multiple candidate power-cap choices yield marginal performance gains that differ only slightly; we quantify this effect by comparing EcoShift with an Oracle in \S~\ref{sec:oracle}.

\begin{table}[t]\label{tab: case_study}
    \centering
    \small
    \setlength{\tabcolsep}{4pt}
    \caption{Performance improvement under a baseline cap of (300 W CPU, 200 W GPU) for two H100 applications, with 200 W of reclaimed power available for redistribution. The table reports the post-allocation CPU/GPU caps selected by each policy under the monotonic upgrade model used in EcoShift.}
    \label{tab:strategy_improvements}
    \begin{tabular}{@{}l l c c@{}}
    \toprule
    \textbf{Policy} & \textbf{App} & \textbf{CPU, GPU Power (W)} & \textbf{Perf. Gain (\%)} \\
    \midrule
    \multirow{2}{*}{EcoShift}
    & raytracing & (300, 300) & 15.57 \\
    & cfd        & (400, 200) & 18.35 \\
    \cmidrule(lr){2-4}
    & \multicolumn{2}{r}{Average} & \textbf{16.96} \\
    \midrule
    \multirow{2}{*}{DPS}
    & raytracing & (350, 250) & 9.61 \\
    & cfd        & (350, 250) & 8.81 \\
    \cmidrule(lr){2-4}
    & \multicolumn{2}{r}{Average} & \textbf{9.21} \\
    \midrule
    \multirow{2}{*}{MixedAdaptive}
    & raytracing & (329, 294) & 17.03 \\
    & cfd        & (354, 221) & 9.29 \\
    \cmidrule(lr){2-4}
    & \multicolumn{2}{r}{Average} & \textbf{13.16} \\
    \bottomrule
    \end{tabular}
\end{table}

\subsection{A Case Study for Detailed Analysis}
We next examine a simple two-application case to show why EcoShift improves average performance. We revisit the example in Figure~\ref{fig:motivation}. Under the baseline caps of 300~W CPU and 200~W GPU, \textit{cfd} benefits strongly from additional CPU power, whereas \textit{raytracing} benefits primarily from additional GPU power.

Table~\ref{tab:strategy_improvements} shows the resulting allocations under the monotonic upgrade model used throughout the paper, where reclaimed power can only raise an application's CPU and GPU caps relative to baseline. EcoShift assigns the full 100~W CPU increase to \textit{cfd}, moving it from $(300,200)$ to $(400,200)$, and assigns the remaining 100~W to \textit{raytracing}, moving it to $(300,300)$. By contrast, DPS applies the same split to both applications, and MixedAdaptive allocates power according to inferred demand rather than predicted marginal gain. The comparison makes the mechanism behind EcoShift clear: it spends power where the next watt is predicted to help most.

Under this policy, EcoShift achieves 16.96\% average improvement, well above DPS (9.21\%) and MixedAdaptive (13.16\%). This case study shows that EcoShift improves average performance not by enforcing a uniform rule, but by matching each power increment to the application's dominant sensitivity within the feasible upgrade set.

\subsection{Gap to Oracle} \label{sec:oracle}
The gap to the Oracle captures the overall suboptimality of EcoShift. This gap comes from two sources: prediction error, because EcoShift optimizes over predicted rather than measured performance surfaces, and optimization error, because the search operates on a discretized formulation. Comparing EcoShift against an exhaustive brute-force Oracle therefore evaluates the full pipeline of prediction plus allocation.

To do this evaluation, we randomly select ten applications from the mixed workload set and repeat this selection process five times. For each selection, we evaluate five different initial power cap configurations, ranging from small to large, and four different reclaimed power budget levels, resulting in a total of 100 test configurations. Each test configuration consists of ten applications with distinct power settings. We limit the evaluation to ten applications per configuration to keep the brute-force search tractable, as larger configurations become prohibitively time-consuming.

\begin{figure}
    \centering
    \includegraphics[width=0.95\linewidth]{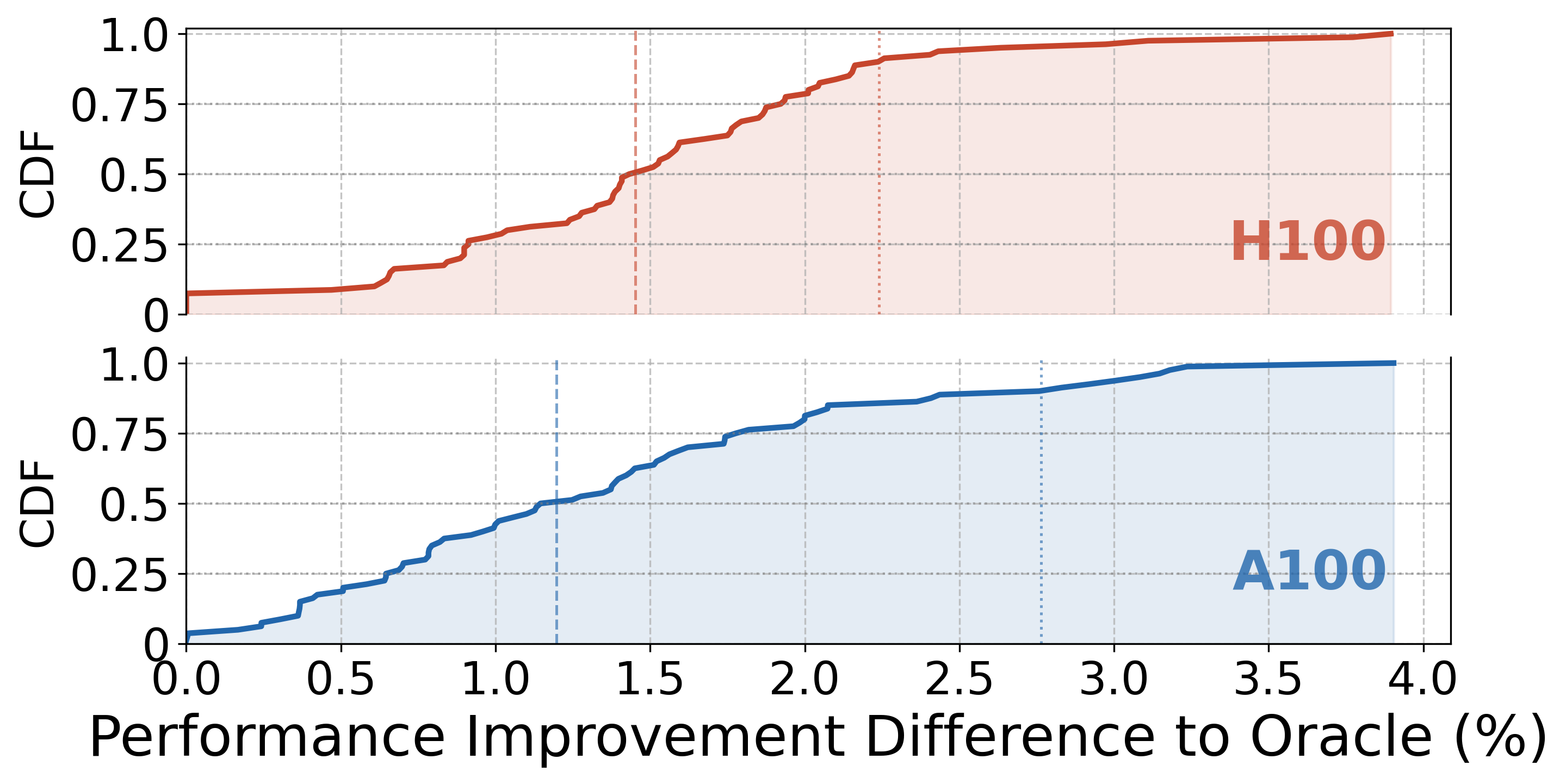}
    \caption{Cumulative distribution function of the performance improvement gap between EcoShift's DP-based search and the Oracle across all 100 test cases. The performance improvement gap is defined as the difference in average performance improvement (\%) between the Oracle, which assumes perfect knowledge of application performance at every power level and identifies the optimal allocation via exhaustive brute-force search, and EcoShift's DP-based search.}
    \label{fig:cdf}
    \vspace{-1.5em}
\end{figure}

Figure~\ref{fig:cdf} plots the cumulative distribution of the performance-improvement gap (in percentage points) between EcoShift's DP-based solution and the Oracle. For example, if the DP-based solution achieves a 10\% performance improvement while the Oracle achieves 15\%, the absolute gap is 5 percentage points. We can observe that, in most cases, the gap between EcoShift and the Oracle remains within a small margin. On H100, EcoShift achieves a median (mean) gap of 1.45 (1.48) percentage points, with a 90th-percentile gap of 2.24 percentage points; 27.5\% and 77.5\% of test cases are within 1 and 2 percentage points of the Oracle, respectively. On A100, EcoShift achieves a median (mean) gap of 1.20 (1.33) percentage points, with a 90th-percentile gap of 2.76 percentage points; 42.5\% and 80.0\% of test cases are within 1 and 2 percentage points of the Oracle, respectively. Overall, on both systems, 90\% of the test cases exhibit less than a 3 percentage-point gap, demonstrating that the DP-based search closely approximates the optimal solution. This shows that the DP-based search remains close to the Oracle while preserving the low-overhead structure needed for online use.

\subsection{Fairness Analysis}
Average performance is the primary objective, but fairness also matters for a cluster-wide policy. We use Jain's fairness index to quantify how evenly performance improvements are distributed across applications. In Equation~\ref{eq:jain}, $x_i$ denotes the performance improvement of application $i$ and $n$ is the number of applications. The index ranges from $1/n$ to $1$, where larger values indicate a more even distribution of gains. Jain's index is appropriate here because it captures relative balance across applications while remaining invariant to scale.

\begin{figure}
    \centering
    \includegraphics[width=1\linewidth]{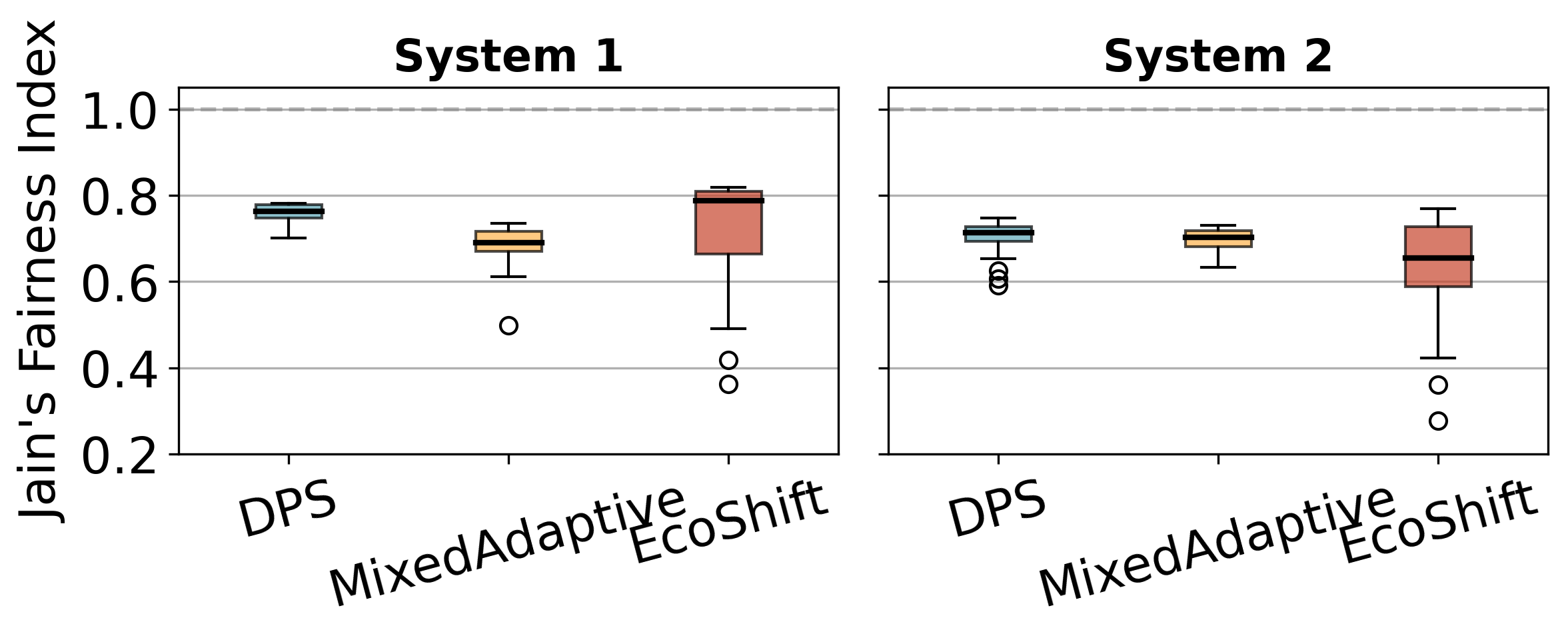}
    \caption{Jain's fairness index of the mixed workloads.}
    \label{fig:fairness}
    \vspace{-1.5em}
\end{figure}

For each experiment configuration, we compute Jain's fairness index over the per-application improvements produced by each policy. Figure~\ref{fig:fairness} summarizes these values for the mixed-workload experiments on both systems. EcoShift exhibits larger variance, which is expected: a policy that targets marginal gains will sometimes concentrate power on a smaller set of highly responsive applications. Even so, EcoShift achieves median fairness comparable to DPS and MixedAdaptive. Thus, EcoShift improves average performance without introducing a systematic fairness collapse.

\section{Discussion}

\textbf{\emph{Key strength.}} A key strength of EcoShift is its ability to make \emph{performance-aware} cluster-wide power distribution decisions for heterogeneous CPU--GPU workloads without expensive offline profiling. By combining lightweight online profiling, an online performance predictor, and a DP-based optimizer, EcoShift can (i) quickly characterize each application’s CPU--GPU power sensitivity and diminishing returns, and (ii) compute a near-optimal reclaimed-power allocation with low enough overhead for practical deployment.

\textbf{\emph{Average-improvement gain.}} Although EcoShift’s improvement over state-of-the-art policies is up to 6\% on average, this metric is computed over the full evaluated workload mix rather than a single best-case application. In an HPC setting, even a few-percent increase in average performance improvement can translate into meaningful throughput benefit across the workload mix.

EcoShift’s gains are not driven by a single outlier. The application-level distributions in Figure~\ref{fig:violin} show that a larger fraction of applications achieve higher improvements under EcoShift than under DPS and MixedAdaptive. While EcoShift may allocate more power to the most power-sensitive applications (potentially reducing improvement for a small subset of less sensitive applications), Figure~\ref{fig:fairness} shows that EcoShift still achieves median fairness comparable to the baselines, indicating that these gains do not come at the cost of severe imbalance.

\textbf{\emph{Online profiling.}} In this work, EcoShift performs online profiling only once for unseen applications to initialize the performance predictor, and then reuses the inferred performance surface for subsequent optimization decisions. This design matches the reality of many HPC applications and deep learning workloads, which often exhibit repetitive phase behavior \cite{online-1} (i.e., recurring power--performance patterns over time). As a result, a short (e.g., one-minute) profiling phase can be sufficient for the predictor to capture the key performance--power behavior and generalize to later phases of the same execution.

\textbf{\emph{Emulation-based evaluation.}} Our emulation-based cluster evaluation enables scalable exploration across many workloads, but it does not capture all effects that may arise in real deployments (e.g., thermal throttling and transient power-cap enforcement dynamics). Strengthening validation with real concurrent executions, reporting end-to-end control overhead (profiling time, prediction latency, and re-optimization frequency), and exploring interactions with scheduler and partitioning mechanisms (e.g., GPU partitioning) would further improve operational relevance.

\textbf{\emph{Evaluation system diversity.}} Our current evaluation focuses on systems equipped with Intel CPUs and NVIDIA GPUs. However, EcoShift’s design is not tied to a specific hardware vendor and can be extended to other heterogeneous architectures that expose similar power-capping mechanisms. This includes emerging APU platforms such as AMD MI300A \cite{mi300a}, which colocate CPU and GPU cores within a single package and operate under a shared socket-level power budget. In these tightly integrated environments, the strong coupling between CPU and GPU resources makes shared memory bandwidth a potentially valuable signal for guiding power distribution decisions.

\section{Related Work}

Power and energy have become first-order constraints in large-scale HPC systems \cite{bergman2008exascale}. To operate within a global power envelope, prior work has explored cluster-wide budgeting, node-level caps, and DVFS-based control \cite{patki2013exploring, google_cloud, doe, offline-1, hybrid-10, offline-10, hybrid-11, online-2, offline-6, dvfs, zheng2025minimizing}. A common baseline is to distribute the available budget uniformly across nodes or applications and then enforce local power caps.

A central limitation of uniform budgeting is that applications respond very differently to power capping. Prior studies show that some workloads cannot fully use their assigned budgets, whereas others experience substantial slowdown under the same cap \cite{ramesh2019understanding}. This observation has motivated a large body of work on redistributing reclaimed power. Existing work can be grouped into three categories: node-level CPU--GPU coordination, application-level redistribution, and cluster-level redistribution.

At the node level, prior work studies how to divide a fixed power budget between CPU and GPU within a single heterogeneous application \cite{ge2016case, aps, aps2}. These methods are valuable for intra-node coordination, but they do not address how reclaimed power should be distributed across multiple applications at the cluster level.

At the application level, prior work focuses on imbalance within a parallel job or among dependent applications. PShifter redistributes power across MPI ranks according to observed load imbalance \cite{gholkar2018pshifter}, while PoDD allocates power across dependent applications to improve the pace of the bottleneck stage \cite{zhang2019podd}. These works optimize performance within an application or a tightly coupled application set rather than across an entire heterogeneous workload mix.

At the cluster level, DPS redistributes reclaimed power according to a fixed-share policy \cite{ding2023dps}, and MixedAdaptive allocates excess power based on demand inferred from current power draw \cite{wilson2021introducing}. These approaches are closest to our setting, but they do not explicitly optimize application-specific marginal gains across both CPU and GPU dimensions. EcoShift differs by using predicted CPU--GPU performance surfaces to allocate reclaimed power where it produces the largest expected marginal gain. To the best of our knowledge, EcoShift is the first cluster-level, performance-aware power distribution framework that jointly considers CPU and GPU while targeting average performance improvement under a fixed reclaimed-power budget.

\section{Conclusion}\label{Conclusion}

As heterogeneous CPU--GPU workloads become more common, cluster-wide power management must account for application-specific sensitivity to CPU and GPU power caps. We presented EcoShift, an online framework that combines lightweight profiling, performance prediction, and dynamic-programming-based optimization to redistribute reclaimed power across applications. Across two heterogeneous platforms and diverse CPU--GPU workloads, EcoShift consistently improves average performance over state-of-the-art policies by up to 6\% while maintaining comparable mean fairness. These results show that performance-aware redistribution can turn reclaimed power into system-level benefit. As future work, we plan to integrate EcoShift with production schedulers such as Slurm, enabling periodic cap updates and re-optimization as applications arrive and depart.

\newpage

\bibliographystyle{ACM-Reference-Format}
\bibliography{bib/sc.bib}

@String{Computing = "Computing" }

@String{Computer = "{IEEE} Computer" }

@String{Springer = "Springer-Verlag" }

@misc{DCGM,
  title = {NVIDIA Data Center GPU Manager},
  author = {NVIDIA},
  howpublished={\url{"https://github.com/NVIDIA/DCGM"}},
  year = {2025},
}

@misc{jain,
  title = {Jain's fairness index},
  author = {},
  howpublished={\url{"https://en.wikipedia.org/wiki/Fairness_measure"}},
  year = {2025},
}

@article{dixit1991spec,
  title={The SPEC benchmarks},
  author={Dixit, Kaivalya M},
  journal={Parallel computing},
  volume={17},
  number={10-11},
  pages={1195--1209},
  year={1991},
  publisher={Elsevier Science Publishers BV Amsterdam, The Netherlands, The Netherlands}
}

@inproceedings{zheng2025minimizing,
  title={Minimizing Power Waste in Heterogenous Computing via Adaptive Uncore Scaling},
  author={Zheng, Zhong and Sultanov, Seyfal and Papka, Michael E and Lan, Zhiling},
  booktitle={Proceedings of the International Conference for High Performance Computing, Networking, Storage and Analysis},
  pages={505--518},
  year={2025}
}

@misc{doe,
  title = {The USDOE Exascale Computing Project–Goals and Challenges},
  author = {Paul Messina},
  howpublished={\url{"https://www.nist.gov/system/files/documents/2017/02/21/messina_nist_20170214.final_.pdf"}},
  year = {2017},
}

@inproceedings{ding2023dps,
  title={DPS: Adaptive Power Management for Overprovisioned Systems},
  author={Ding, Jianru and Hoffmann, Henry},
  booktitle={Proceedings of the International Conference for High Performance Computing, Networking, Storage and Analysis},
  pages={1--14},
  year={2023}
}

@inproceedings{hec,
  title={A benchmark suite for improving performance portability of the sycl programming model},
  author={Jin, Zheming and Vetter, Jeffrey S},
  booktitle={2023 IEEE International Symposium on Performance Analysis of Systems and Software (ISPASS)},
  pages={325--327},
  year={2023},
  organization={IEEE}
}

@inproceedings{david2010rapl,
  title={RAPL: Memory power estimation and capping},
  author={David, Howard and Gorbatov, Eugene and Hanebutte, Ulf R and Khanna, Rahul and Le, Christian},
  booktitle={Proceedings of the 16th ACM/IEEE international symposium on Low power electronics and design},
  pages={189--194},
  year={2010}
}

@misc{NVML,
  author = {Nvidia},
  title ={{NVML}},
  howpublished={\url{"https://developer.nvidia.com/management-library-nvml"}},
  year = {2025}
}

@misc{mi300a,
  author = {AMD},
  title ={{AMD MI300A}},
  howpublished={\url{"https://www.amd.com/en/products/accelerators/instinct/mi300/mi300a.html"}},
  year = {2025}
}

@inproceedings{hu2020altis,
  title={Altis: Modernizing gpgpu benchmarks},
  author={Hu, Bodun and Rossbach, Christopher J},
  booktitle={2020 IEEE International Symposium on Performance Analysis of Systems and Software (ISPASS)},
  pages={1--11},
  year={2020},
  organization={IEEE}
}

@inproceedings{patki2013exploring,
  title={Exploring hardware overprovisioning in power-constrained, high performance computing},
  author={Patki, Tapasya and Lowenthal, David K and Rountree, Barry and Schulz, Martin and De Supinski, Bronis R},
  booktitle={Proceedings of the 27th international ACM conference on International conference on supercomputing},
  pages={173--182},
  year={2013}
}

@inproceedings{google_cloud,
  title={Data center power oversubscription with a medium voltage power plane and priority-aware capping},
  author={Sakalkar, Varun and Kontorinis, Vasileios and Landhuis, David and Li, Shaohong and De Ronde, Darren and Blooming, Thomas and Ramesh, Anand and Kennedy, James and Malone, Christopher and Clidaras, Jimmy and others},
  booktitle={Proceedings of the Twenty-Fifth International Conference on Architectural Support for Programming Languages and Operating Systems},
  pages={497--511},
  year={2020}
}

@inproceedings{yoo2003slurm,
  title={Slurm: Simple linux utility for resource management},
  author={Yoo, Andy B and Jette, Morris A and Grondona, Mark},
  booktitle={Workshop on job scheduling strategies for parallel processing},
  pages={44--60},
  year={2003},
  organization={Springer}
}

@inproceedings{gholkar2018pshifter,
  title={Pshifter: Feedback-based dynamic power shifting within hpc jobs for performance},
  author={Gholkar, Neha and Mueller, Frank and Rountree, Barry and Marathe, Aniruddha},
  booktitle={Proceedings of the 27th International Symposium on High-Performance Parallel and Distributed Computing},
  pages={106--117},
  year={2018}
}

@inproceedings{ramesh2019understanding,
  title={Understanding the impact of dynamic power capping on application progress},
  author={Ramesh, Srinivasan and Perarnau, Swann and Bhalachandra, Sridutt and Malony, Allen D and Beckman, Pete},
  booktitle={2019 IEEE International Parallel and Distributed Processing Symposium (IPDPS)},
  pages={793--804},
  year={2019},
  organization={IEEE}
}

@inproceedings{zhang2019podd,
  title={PoDD: power-capping dependent distributed applications},
  author={Zhang, Huazhe and Hoffmann, Henry},
  booktitle={Proceedings of the International Conference for High Performance Computing, Networking, Storage and Analysis},
  pages={1--23},
  year={2019}
}

@article{khan2018rapl,
  title={Rapl in action: Experiences in using rapl for power measurements},
  author={Khan, Kashif Nizam and Hirki, Mikael and Niemi, Tapio and Nurminen, Jukka K and Ou, Zhonghong},
  journal={ACM Transactions on Modeling and Performance Evaluation of Computing Systems (TOMPECS)},
  volume={3},
  number={2},
  pages={1--26},
  year={2018},
  publisher={ACM New York, NY, USA}
}

@misc{perf,
  author = {Linux},
  title ={{perf — Linux manual page}},
  howpublished={\url{"https://man7.org/linux/man-pages/man1/perf.1.html"}},
  year = {2024}
}

@article{atomic2013lammps,
  title={Lammps},
  author={Atomic, Large-scale and Simulator, Molecular Massively Parallel},
  journal={available at: http:/lammps. sandia. gov},
  year={2013}
}

@article{van2005gromacs,
  title={GROMACS: fast, flexible, and free},
  author={Van Der Spoel, David and Lindahl, Erik and Hess, Berk and Groenhof, Gerrit and Mark, Alan E and Berendsen, Herman JC},
  journal={Journal of computational chemistry},
  volume={26},
  number={16},
  pages={1701--1718},
  year={2005},
  publisher={Wiley Online Library}
}

@article{lefurgy2008power,
  title={Power capping: a prelude to power shifting},
  author={Lefurgy, Charles and Wang, Xiaorui and Ware, Malcolm},
  journal={Cluster Computing},
  volume={11},
  pages={183--195},
  year={2008},
  publisher={Springer}
}

@inproceedings{wilson2021introducing,
  title={Introducing application awareness into a unified power management stack},
  author={Wilson, Daniel C and Jana, Siddhartha and Marathe, Aniruddha and Brink, Stephanie and Cantalupo, Christopher M and Guttman, Diana R and Geltz, Brad and Lawson, Lowren H and Al-Rawi, Asma H and Mohammad, Ali and others},
  booktitle={2021 IEEE International Parallel and Distributed Processing Symposium (IPDPS)},
  pages={320--329},
  year={2021},
  organization={IEEE}
}

@inproceedings{dvfs,
  title={Memory power management via dynamic voltage/frequency scaling},
  author={David, Howard and Fallin, Chris and Gorbatov, Eugene and Hanebutte, Ulf R and Mutlu, Onur},
  booktitle={Proceedings of the 8th ACM international conference on Autonomic computing},
  pages={31--40},
  year={2011}
}

@misc{ecp,
  title ={{ECP proxy apps suite}},
  howpublished={https://proxyapps.exascaleproject.org/ ecp- proxy- apps- suite/.},
  year={2025}
}

@inproceedings{farrell2021mlperf,
  title={MLPerf™ HPC: A holistic benchmark suite for scientific machine learning on HPC systems},
  author={Farrell, Steven and Emani, Murali and Balma, Jacob and Drescher, Lukas and Drozd, Aleksandr and Fink, Andreas and Fox, Geoffrey and Kanter, David and Kurth, Thorsten and Mattson, Peter and others},
  booktitle={2021 IEEE/ACM Workshop on Machine Learning in High Performance Computing Environments (MLHPC)},
  pages={33--45},
  year={2021},
  organization={IEEE}
}

@article{bergman2008exascale,
  title={Exascale computing study: Technology challenges in achieving exascale systems},
  author={Bergman, Keren and Borkar, Shekhar and Campbell, Dan and Carlson, William and Dally, William and Denneau, Monty and Franzon, Paul and Harrod, William and Hill, Kerry and Hiller, Jon and others},
  journal={Defense Advanced Research Projects Agency Information Processing Techniques Office (DARPA IPTO), Tech. Rep},
  volume={15},
  pages={181},
  year={2008}
}

@inproceedings{offline-1,
  title={GPU power prediction via ensemble machine learning for DVFS space exploration},
  author={Dutta, Bishwajit and Adhinarayanan, Vignesh and Feng, Wu-chun},
  booktitle={Proceedings of the 15th ACM International Conference on Computing Frontiers},
  pages={240--243},
  year={2018}
}

@inproceedings{offline-6,
  title={GPGPU power modeling for multi-domain voltage-frequency scaling},
  author={Guerreiro, Joao and Ilic, Aleksandar and Roma, Nuno and Tomas, Pedro},
  booktitle={2018 IEEE International Symposium on High Performance Computer Architecture (HPCA)},
  pages={789--800},
  year={2018},
  organization={IEEE}
}

@article{offline-10,
  title={GPGPU performance estimation with core and memory frequency scaling},
  author={Wang, Qiang and Chu, Xiaowen},
  journal={IEEE Transactions on Parallel and Distributed Systems},
  volume={31},
  number={12},
  pages={2865--2881},
  year={2020},
  publisher={IEEE}
}

@inproceedings{hybrid-10,
  title={Predictable GPUs Frequency Scaling for Energy and Performance},
  author={Fan, Kaijie and Cosenza, Biagio and Juurlink, Ben},
  booktitle={Proceedings of the 48th International Conference on Parallel Processing},
  pages={1--10},
  year={2019},
  publisher={ACM}
}

@article{hybrid-11,
  title={DVFS-aware application classification to improve GPGPUs energy efficiency},
  author={Guerreiro, João and Ilic, Aleksandar and Roma, Nuno and Tomás, Pedro},
  journal={Parallel Computing},
  volume={83},
  pages={93--117},
  year={2019},
  publisher={Elsevier}
}

@inproceedings{online-1,
  title={Indicator-directed dynamic power management for iterative workloads on GPU-accelerated systems},
  author={Zou, Pengfei and Li, Ang and Barker, Kevin and Ge, Rong},
  booktitle={2020 20th IEEE/ACM International Symposium on Cluster, Cloud and Internet Computing (CCGRID)},
  pages={559--568},
  year={2020},
  organization={IEEE}
}

@inproceedings{online-2,
  title={Improving gpu energy efficiency through an application-transparent frequency scaling policy with performance assurance},
  author={Zhang, Yijia and Wang, Qiang and Lin, Zhe and Xu, Pengxiang and Wang, Bingqiang},
  booktitle={Proceedings of the Nineteenth European Conference on Computer Systems},
  pages={769--785},
  year={2024}
}

@article{cornelius2025extracting,
  title={Extracting Practical, Actionable Energy Insights from Supercomputer Telemetry and Logs},
  author={Cornelius, Melanie and Cross, Greg and Shilpika, Shilpika and Dearing, Matthew T and Lan, Zhiling},
  journal={arXiv preprint arXiv:2505.14796},
  year={2025}
}

@inproceedings{hong2010integrated,
  title={An integrated GPU power and performance model},
  author={Hong, Sunpyo and Kim, Hyesoon},
  booktitle={Proceedings of the 37th annual international symposium on Computer architecture},
  pages={280--289},
  year={2010}
}

@inproceedings{lee2011improving,
  title={Improving throughput of power-constrained GPUs using dynamic voltage/frequency and core scaling},
  author={Lee, Jungseob and Sathisha, Vijay and Schulte, Michael and Compton, Katherine and Kim, Nam Sung},
  booktitle={2011 International Conference on Parallel Architectures and Compilation Techniques},
  pages={111--120},
  year={2011},
  organization={IEEE}
}

@article{aps,
  title={Adaptive power shifting for power-constrained heterogeneous systems},
  author={Ortega, Cristobal and Alvarez, Lluc and Buyuktosunoglu, Alper and Bertran, Ramon and Rosedahl, Todd and Bose, Pradip and Moreto, Miquel},
  journal={IEEE Transactions on Computers},
  volume={72},
  number={3},
  pages={627--640},
  year={2022},
  publisher={IEEE}
}

@article{aps2,
  title={Dynamic Power Management Through Multi-agent Deep Reinforcement Learning for Heterogeneous Systems},
  author={Wang, Yiming and Zhang, Weizhe and Hao, Meng and Kong, Weizhi and Wen, Yuan},
  journal={ACM Transactions on Architecture and Code Optimization},
  year={2025},
  publisher={ACM New York, NY}
}

@inproceedings{ge2016case,
  title={The case for cross-component power coordination on power bounded systems},
  author={Ge, Rong and Feng, Xizhou and He, Yangyang and Zou, Pengfei},
  booktitle={2016 45th International Conference on Parallel Processing (ICPP)},
  pages={516--525},
  year={2016},
  organization={IEEE}
}

@inproceedings{srivastava2022penelope,
  title={Penelope: peer-to-peer power management},
  author={Srivastava, Tapan and Zhang, Huazhe and Hoffmann, Henry},
  booktitle={Proceedings of the 51st International Conference on Parallel Processing},
  pages={1--11},
  year={2022}
}

@article{zheng2025coordinated,
  title={Coordinated power management on heterogeneous systems},
  author={Zheng, Zhong and Lan, Zhiling and Wu, Xingfu and Taylor, Valerie E and Papka, Michael E},
  journal={arXiv preprint arXiv:2508.07605},
  year={2025}
}

\end{document}